\newenvironment{namelist}[1]{%
\begin{list}{}
    {
      
      \settowidth{\labelwidth}{#1}
      \setlength{\leftmargin}{1.1\labelwidth}
    }
  }{%
\end{list}}

\documentstyle[times,
twoside,csit,epsf]{article}

\author{\large V.E.Wolfengagen \vspace{1.52mm} \\
{\normalsize Vorotnikovsky per, 7, bld. 4} \\
{\normalsize Dept. for Advanced Computer Studies and Information Technologies}\\
{\normalsize Institute for Contemporary Education ``JurInfoR-MSU''} \\
{\normalsize Moscow, 103006, Russia} \\
{\normalsize e-mail: {\tt vew@jmsuice.msk.ru}}
    \date{}
    }
\title{Logic, Individuals and Concepts
}

\newtheorem{df}{\sf Definition}[section]

\institution{}

\begin{document}

\setcounter{page}{141}
\bibliographystyle{alpha}

\markboth{Logic, Individuals and Concepts}
{Workshop on Computer Science and Information Technologies CSIT'2000,
Ufa, Russia, 2000}

\maketitle


\begin{abstract}
\noindent This extended abstract gives a
brief outline of the connections
between {\em the descriptions} and {\em variable concepts}.
Thus, the notion of a concept is extended to include both
the syntax and semantics features. The evaluation map in use
is parameterized by a kind of computational environment, the
{\em index}, giving rise to indexed concepts.
The concepts are
inhabited into language by the {\em descriptions}
from the higher order logic.
In general the idea of object-as-functor
should assist the designer to
outline a programming tool in {\em conceptual shell} style.
\end{abstract}


\section*{Introduction}
\addcontentsline{toc}{section}{Introduction}

The notion of an object arises for different purposes,
especially in
specific applied systems and often inspired by accidents.
Issue where the objects come from usually is distinct from
exact development and tends to mathematical consideration.
The remarks here can be taken as a suggestion to group
numerous aspects of `object' to result in a general computational
framework that gives a suitable scheme. This scheme can be useful
as a {\em primitive frame} to put important ideas of data objects
modeling in a certain order.

The individuals in a problem domain
are briefly discussed in Section~\ref{str-prob-dom},
and they are thought to be
coupled into a single set $D$.
Most of them are {\em possible} objects
and can be converted into the {\em actual} objects.
The difference is
captured by the {\em assignments} that play a role of
possible worlds, as shown in Section~\ref{log-ind-conc}.
The idea of possible worlds is more understood in a theoretical
research and is not yet adopted as everyday tool of database
designer and developer. For convenience imagine the actual object
as an assigned possible object with respect to an index being
specified.
In a theory the set of possible worlds is represented by a suitable
mathematical structure e.g. a category. To capture the dynamics
of a problem domain this category represents the `evolving of
events', that is shortly covered in Section~\ref{var-dom-var-conc}.

The descriptions of selfcontained couples of individuals generate
concepts. Concepts are described and represented according to
scheme of {\em comprehension} in a higher order logic.



\section{Structuring a problem domain}\label{str-prob-dom}

The notion of a {\em problem domain} is a cross point
for a lot of researches.  A choice of the
{\em object} brings the most difficulties
 -- this is an atomic entity, the unit that generates
the compound entities. Thus the objects induce an inductive class
that contains all their representations.

Objects do not produce the homogeneous set. It is separated into
at least three counterparts: sets of the
{\em actual} $A$, {\em possible} $D$ and {\em virtual} $V$ objects.
As usually they induce the natural inclusion
$A\subseteq D\subseteq V$. In fact, for $i\in I$ this inclusion
transforms into $A_i\subseteq D\subseteq V$.


\subsection{Objects as individuals}

The {\em individuals} in a problem domain take part in all the
constructions. Mainly the individuals are fixed by the
individualizing functions. The discussions of their nature is
excessively troublesome. But the most convenient reason is the
possibility to express the individuals within the framework of
some {\em logical language}. Therefore the properties of this
language have the dominant importance. A selectivity of the
language has to enable capturing those {\em minimal} objects named
individuals. Furthermore the individuals have to be comprehensed
to give rise to more vivid objects. Comprehension in a language
takes the central place.

\subsection{Comprehension}
Let $\Phi$ be an individualizing function that corresponds
to the logical sentence or formula. The selective power of
this formula permits the unique identification of a
distinct object $\overline{h^{\prime}}$ in a language. \\
This uniqueness is announced as a principle of comprehension
(with {\em the description} {$\cal I$}):
\begin{center}
$
\parallel {\cal I}y\Phi (y)\parallel =h \Leftrightarrow
\{h\}=\{h^{\prime}\in D \mid~\parallel\Phi (\overline{h^{\prime}}) \parallel
=~1\}$ \\
$T\equiv\{h:D\mid\Phi\}\equiv{\cal I}y:\lbrack D\rbrack\forall
h:D (\Phi \leftrightarrow y(h)) $
\end{center}
Second expression enables type $T$ within a language, and the type
indicates the set of individuals for a {\em power sort} $[D]$.


\section{Logic, individuals and concepts}\label{log-ind-conc}

Linkage of individuals and formulae is established in
a logical language.
\subsection{Language}
The {\em language} tends to define and manipulate the objects
of different kinds and gives a logical {\em snapshot}. This snapshot
does not depend on any external parameters the same way as in
a database theory: \\
{\small
\indent Object::=Atom $\mid$ Complex \\
\indent Atom::=Constant $\mid$ Variable \\
\indent Complex::=Constant\_function(Object)$\mid$\\
\indent\indent $\mid$\lbrack Object,Object \rbrack \\
\indent\indent $\mid$ Object(Object) \\
\indent\indent $\mid$ Object $\in$ Object \\
\indent Logical\_formula::=Equation $\mid $ Compound \\
\indent\indent Equation::=(Variable=Complex) \\
\indent\indent Compound::=Logical\_formula $\wedge$ Logical\_formula $\mid$\\
\indent\indent\indent $\mid$ Logical\_formula $\vee$ Logical\_formula \\
\indent\indent\indent $\mid$ Logical\_formula $\Rightarrow$ Logical\_formula \\
\indent\indent\indent $\mid$ $\exists$Variable.Logical\_formula \\
\indent\indent\indent $\mid$ $\forall$Variable.Logical\_formula
} \\
The objects with the proper computational aspects are, as usually,
pairs $\lbrack\cdot , \cdot\rbrack$, applications $\cdot(\cdot )$,
and inclusions $\cdot\in \cdot$ which generate a class of equations.
The equations are counterparts in the compounds. The target of this is to
support the object of a special kind, namely the concept.

\subsection{Building the concepts}

Concepts are mainly the basic building blocks. They are separated into
generic concepts and indexed concepts.
Note that new concepts generate the {\em definitional dimension} and are
introduced by the descriptions like:
\[
\begin{array}{lcl}
New\_concept &=& {\cal I}y:Power\_sort \\
             & & \forall x:Sort \\
             & &( y(x) \leftrightarrow Logical\_formula(x) )
\end{array}
\]

\index{Object!intentional} {\em Generic concepts} are often used as a
kind of the {\em representation}. They are intentional objects -- and are
correspondent to {\em sort} or {\em type} symbols -- which are interpreted
as sets. Initially the generic concepts are established to represent
generic ideas of physical
or abstract objects that are distinct and understood in a problem domain.


\subsubsection{Indexed concept}

\indent {\em Indexed concept} results in a family of concepts
each of them being dependent on the parameter (index, assignment etc.).
Its counterparts are as follows:
\begin{namelist}{{\tt \makebox[0.60in][l]
{$\parallel\Phi(h)\parallel i $}} {\it }}

\item[{\tt \makebox[0.60in][l]{$i\in I$}}{}] {\it Index:}
Object $i$ is selected from a set $I$. Index identifies the valid
configuration of database.
\item[{\tt \makebox[0.60in][l]{$h(i)\in T$}}{}] {\it Type:}
Object $h(i)$ is an indexed individual that is contained in the type.
\item[{\tt \makebox[0.60in][l]
{$\parallel\Phi(\overline{h})\parallel i $}}{}] {\it Substitution:}
Object $\overline{h}$ with type $I\times T$ is a valid substitution of
formula $\Phi$ and $h$ is isomorphic to pair $\lbrack i, h(i)\rbrack$
for projections $p, q$: $p(\lbrack i, h(i)\rbrack)=i$ and
$q(\lbrack i, h(i)\rbrack)=h(i)$.
\item[{\tt \makebox[0.60in][l]
{$C'(\{i\})$}}{}] {\it Instantiation:} Object
$C'(\{i\})=\{h(i)\mid~\parallel\Phi(\overline{h})\parallel i =1\}$
is an indexed concept and $C'(\{i\})\in \lbrack T \rbrack$. It is
the intentional object with the extension that is generated by the
substitutions of $\Phi$.
\item[{\tt \makebox[0.60in][l]
{$C(I)$}}{}] {\it Variable concept:}
Object $C(I)\in \lbrack I \times T \rbrack$ is a variable concept
that generates the family
of indexed concepts:
$C(I)=\{C(\{i\})\}_{i\in I}=
      \{h(i)\mid~\parallel\Phi(\overline{h})\parallel i =1\}_{i\in I}$.
\end{namelist}

Therefore indexed concept indicates the elements of $\lbrack T\rbrack$:
 when $i$ ranges $I$ then $C(\{i\})$ ranges $\lbrack T\rbrack$.


\subsubsection{Evolvent}

Let $B$ and $I$ be the sets of indexes.
The mapping $f:B \rightarrow I$ reaches elements $B$ from $I$ so
that $f(b)=i$ for some $i\in I, b\in B$. If the elements of $B$ and
$I$ are understood as events then $f$ is therefore assumed to be
evolvent of events. The reversed order of reading the mapping $f$
is selected for technical convenience.

Any way, evolvents capture more dynamics in a problem domain. The
steps for indexing concept as above are following.

\begin{namelist}{{\tt \makebox[0.65in][l]
{$\parallel\Phi(h)\parallel_f b $}} {\it }}

\item[{\tt \makebox[0.65in][l]{$b\in B$}}{}] {\it Index:}
Object $b$ is selected from a set $B$. Index identifies the `new'
configuration of database.
\item[{\tt \makebox[0.65in][l]{$h(b)\in T$}}{}] {\it Type:}
Object $h(b)$ is an indexed individual that type contains. For evolvent
$f$ the composition gives $h(i)=(h\circ f)(b)$. Thus $b$ is the `new'
configuration of database and $i$ is an `old' one. From this point
of view $h\circ f$ is an $f$-shifted image of individual $h$ along
the evolvent $f$. Therefore  the
possibility to observe the $f$-shifted individuals $h\circ f$ in $b$
instead of $h$-individuals in $i$ is available. The remainder of steps
repeats the steps above with slightly modified indexes.
\item[{\tt \makebox[0.65in][l]
{$\parallel\Phi(\overline{h})\parallel_f b $}}{}] {\it Substitution:}
Object $\overline{h}\circ f$ with type $B\times T$ is a valid substitution of
formula $\Phi$ and $h$ is isomorphic to pair $\lbrack b, (h\circ f)(b)\rbrack$
for projections $p, q$: $p(\lbrack b, (h\circ f)(b)\rbrack)=b$ and
$q(\lbrack b, (h\circ f)(b)\rbrack)=h(i)$.
\item[{\tt \makebox[0.65in][l]
{$C'_f(b)$}}{}] {\it Instantiation:} Object $C'_f(\{b\})=
\{(h\circ f)(b)\mid~\parallel\Phi(\overline{h})\parallel_f b =1\}$
is an indexed concept and $C'_f(\{b\})\in \lbrack T \rbrack$. It
is the intentional object with the extension that is generated by
the substitutions of $\Phi$.
\item[{\tt \makebox[0.65in][l]
{$C_f(B)$}}{}] {\it f-Concept:}
Object $C_f(B)\in \lbrack B \times T \rbrack$ is a variable concept
that generates the family
of indexed concepts:
$C_f(B)=\{C_f(\{b\})\}_{b\in B}=
      \{(h\circ f)(b)\mid~
         \parallel\Phi(\overline{h})\parallel_f b =1\}_{b\in B}$.
\item[{\tt \makebox[0.65in][l]
{$C(B)$}}{}] {\it Concept:}
Object $C_f(B)$ is a subset of $C(B)$ for the clear reason: set of
configurations $B$ accepts only those events that are $f$-shifted
from $I$ and $C_f(B)\subseteq C(B)$.
\end{namelist}
(Note that in combinatory logic the substitution has a slightly
modified form:
$\parallel (\lambda x.\Phi)\overline{h}\parallel i =\lbrack
h(i)/x\rbrack~(\parallel \Phi \parallel i) $.)


\subsection{Computational model}

\indent Language has to be enforced by the external parameters, or
stages of knowledge, or assignments etc. Assignments enable language
to capture a family of snapshots, or {\em view}. This view is partially
analogous to the view in a database theory, but only partially. The
computational model with views is the following. \\
\[
\begin{array}{lcl}
Concept&=&individual^{assignment} \\
 individual&=&{state}^{assignment} \\
 individual&=&Concept(assignment) \\
 state&=&individual(assignment) \\
 Logic &=&
      Logical\_formula^{assignment}  \\
       & & | Truth\_value^{assignment} \\
 Logical\_formula&=&Logic(assignment)  \\
                 &=&Truth\_value \\
    Truth\_value&=&\{true,false\} \\
\end{array}
\]
To explicate the advantages of the approach let us first establish
the links between logic and concepts.


\subsection{Logical revelation of the concepts}

Logical formulae generate the concepts by the
set-theoretic definitions:

$Concept =\{individual\mid Logical\_formula\} $

The concepts are fixed in a language by the descriptions with
the comprehension captured from a higher order logic. Think of
objects as having being described.


To manipulate {\em objects} the computational tool is needed.
The basic set of the objects is the following: $\perp$ (the logical
constant), $g$ (the functional constants), $(\cdot(\cdot))$ (the
application operator, or the functional variable), $ \in  $
(the set constructor).

The {\em model} is to be constructed to reflect the resulting
evaluations by the induction on the object complexity:
\begin{center}
Objects:
\end{center}
\[
\begin{array}{lcl}
\parallel \perp \parallel I
  & = & false \\
\parallel x=y \parallel I
  & = & \parallel x \parallel I = \parallel y \parallel I \\
\parallel gx \parallel I
  & = & g\circ \parallel x \parallel I \\
\parallel \lbrack x,y \rbrack \parallel I
  & = & \lbrack \parallel x \parallel I~,~ \parallel y \parallel I \rbrack \\
\parallel x(y) \parallel I
  & = & (\parallel x \parallel_{{\bf 1}_A} I) (\parallel y \parallel I) \\
\parallel y \in x \parallel I
  & = & \parallel y \parallel I \in \parallel x \parallel_{{\bf 1}_A} I
\end{array}
\]
\begin{center}
Logical objects:
\end{center}
\[
\begin{array}{lcl}
\parallel (\Phi \land \Psi) \parallel I
     & = & \parallel \Phi \parallel I~\land~ \parallel \Psi \parallel I \\
\parallel (\Phi \lor \Psi) \parallel I
     & = & \parallel \Phi \parallel I~\lor~ \parallel \Psi \parallel I \\
\parallel (\Phi \Rightarrow \Psi) \parallel I
     & = & f:B \rightarrow I ~\&
        ~ \parallel \Phi \parallel_f B ~\Rightarrow~\\
     && ~\Rightarrow~ \parallel \Psi \parallel_f B \\
\parallel \forall x.\Phi \parallel I
     & = & f:B \rightarrow I ~\&
       ~ b \in H_T(B) ~\Rightarrow~\\
     & & ~\Rightarrow~ \lbrack b/x \rbrack \parallel \Phi \parallel_f B \\
\parallel \exists x.\Phi \parallel I
     & = & \exists a \in H_T(I). \lbrack a/x \rbrack \parallel \Phi \parallel I
\end{array}
\]

The computational model above gives rise to the syntax-and-semantic object
$\parallel\cdot\parallel\cdot$, the evaluation map. To understand
its properties the idea of variable domain is needed. Note that the
construction separates the system of concepts and the managing of them.



In the above the notation $\lbrack a/x \rbrack (\parallel \cdot \parallel
\cdot )$ means the valuation $\parallel \cdot \parallel \cdot$ as fixed so
that $a$ matches $x$. The notation $\parallel \cdot \parallel_f \cdot$ means
the valuation that matches $\parallel x \parallel_f$ with each of the
relevant variables $x$ for $f$-shifted valuation.

\section{Variable domains and
          variable concepts}\label{var-dom-var-conc}


\subsection{Variable domains}

Suppose that all the {\em individuals} are gathered into a single
general set $D$ -- a set of all {\em possible} individuals. They
are possible relative to some predefined theory, and this assumption
is not too restrictive but is fruitful to bring more accuracy into
reasonings. Assume the {\em events} concerning the management of the
{\em living conditions} for individuals. The additional assumption needs
the
\index{Domain!variable} law of `event evolving', say
$f:B\rightarrow I$ when the events evolve {\em from} $I$ {\em to} $B$
(pay attention to the reversed order). The set $I$ enables some
{\em possible world} for the individuals, and so does $B$. The notion
of possible world is more known in a database theory as the
{\em valid configuration} of database.

Thus the individuals depend on $I$ and are mainly the
functions like $h:I\rightarrow T$ where $T$ is a {\em type} symbol
that indicates the distinct couple of individuals. The different way
to understand the individuals leads to the {\em pairs}
$h^{\prime }=[i,h(i)]$ where $i \in I$ and $hi \in T$.

\begin{df}[variable domain]
The {\em variable domain} arises as the set
$H_T(I)=\{h\mid h:I\rightarrow T\}$ giving rise to the
mathematical object that is known as the functor $H_T(I)$
with the parameters $T$ and $I$. When $i$ ranges  $I$ then $h(i)$
ranges $T$.
\end{df}
Indeed, the variable domain varies along the evolvent when the
events evolve.  For $f:B\rightarrow I$ the functor $H_T(f)$ represents
the transition from the `old' world $I$ to the `new' world $B$:
$H_T(f): H_T(I) \rightarrow H_T(B)$. The reasoning in terms of elements
gives the mapping
$H_T(f):h\in H_T(I)\mapsto h\circ f\in H_T(B)$
for evolvent $f:B\rightarrow I$, `new' variable domain
$ H_T(B)=\{h'\mid h':B\rightarrow T\}$
and $H_T(f)(h)=h'=h\circ f$.
All of this establishes a variable domain as
the {\em contravariant} functor and exemplifies an operation
of $f$-shifting. \\

\subsection{Variable concepts}

The variable domain denotes the range of a variable.
The logical formula $\Phi(x)$ with free variable $x\in H_T(I)$
reflects the idea. The logic and problem domain are separated
by the {\em evaluation}
$ \parallel \cdot \parallel \cdot $ that depends both on
the formula $\Phi(x)$ under evaluation and the world $I$ for individuals.
As an object the valuation has the properties of functor: \\
\[
\parallel \cdot \parallel I\ {\rm for}\
f:B \rightarrow I\ {\rm is}\ \parallel \cdot \parallel_f B
\]
where $\parallel \cdot \parallel_f \cdot$ is the $f$-shifted
valuation of some formula. \\
To build a {\em subset} of $H_T(I)$ the set
$\{h\mid~ \parallel \Phi(x) \parallel I=1 \}$
should be fixed, denoted by $C(I)$ and named a {\em concept}.
Note that this kind of concept depends on the world $I$:
\[ C(I)=\{h\mid~ \parallel \Phi(x) \parallel I=1 \}\subseteq H_T(I) \]

Note that $C(\{i\} ) \subseteq T \in \lbrack T \rbrack$ for $i\in I$,
type $T$ and power type $\lbrack T\rbrack$.
For future needs arrange the used terms:
$i$ -- the {\em assignment},
$I$ -- the {\em set} of assignments,
$T$ -- the {\em type};
\[
\begin{array}{c}
Type{\_}of(i)=I,\  Type{\_}of(h(i))=T,    \\
Type{\_}of(\lbrack i , h(i) \rbrack )=I \times T,  \\
Type{\_}of(h)=I \rightarrow T,\
Instance{\_}of(i)=I,    \\
i \in I,\
h(i) \in T,\
h=\lbrack i , h(i) \rbrack .
\end{array}
\]


\subsection{Diagrams}

\begin{figure*}
\centerline{\epsfbox{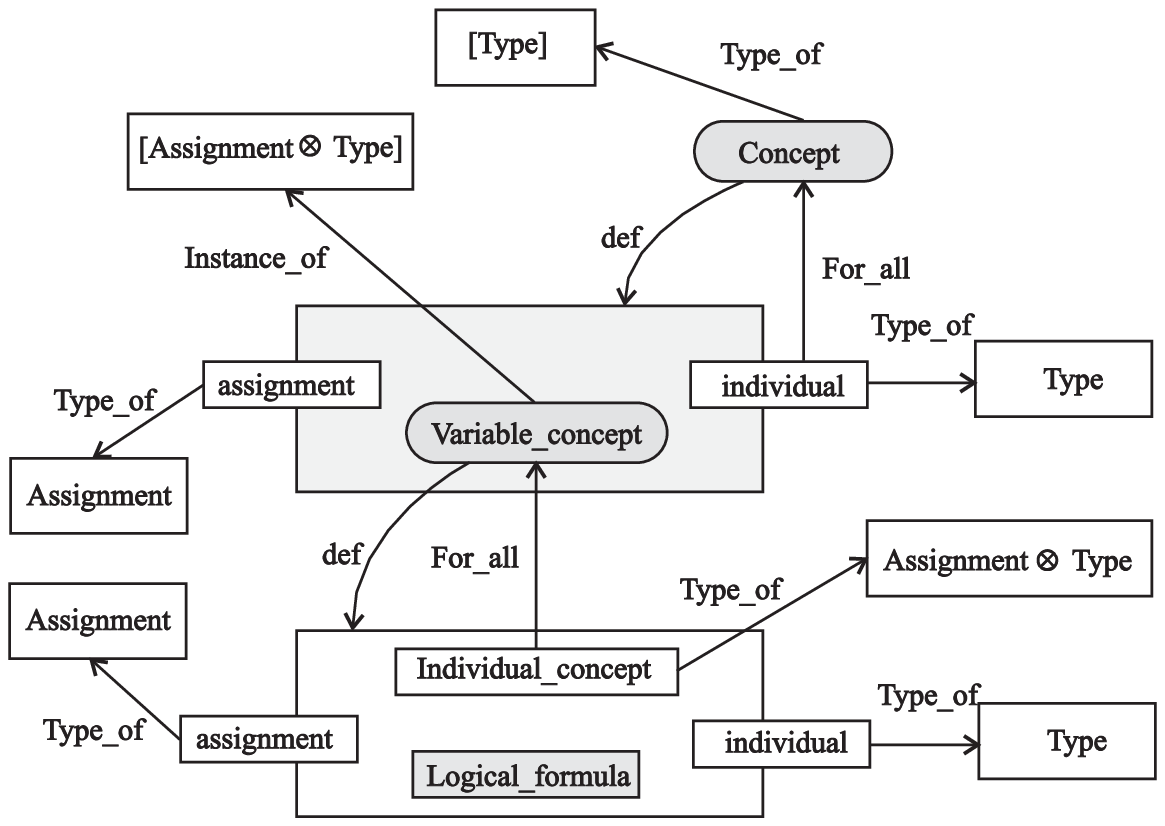}} \caption{Diagram of variable
concept.}\label{fig:1}
\end{figure*}

The sample diagram for all data objects in use
is shown in Figure~\ref{fig:1}.

The entities in this figure conform to the following set
of domain equations:\\
Type\_of(assignment)=Assignment \\
Type\_of(individual)=Type \\
Type\_of(individual\_concept)=Assignment $\times$ Type \\
Type\_of(individual\_concept)=Variable\_concept \\
Type\_of(Variable\_concept)= \\
    \hspace*{1in}      = Power\_set(Assignment $\times$ Type) \\
Type\_of(individual)=Concept \\
Type\_of(Concept)=Power\_set(Type)

\section*{Conclusions}
\addcontentsline{toc}{section}{Conclusions}

The connection of the variable domains and variable concepts is
left out of this extended abstract scope. The typeless
computational model based on variable domains is left out of this
extended abstract scope as well. As could be shown, this general
model, free of any choice of specified evaluation map, does exist
corresponding to an equational theory with products.

Some of the important {\em data objects} such as {\em type},
{\em relation}, {\em function value} and {\em abstraction}
can be derived from the comprehension.

The objects in use inherit both syntax and semantic of the
initial idea of object. This leads to and object-as-functor
computations and generates a family of variable domains.

Variable concepts can be embedded into the computational model
and inherit the logical properties of the objects. The higher order
logic (with {\em the descriptions}) is in use.

Variable concept gives a natural representation of database {\em view}.
Indexed concepts give a sound basis to create database {\em snapshots}.
The parameter of variable concept ranges a category of evolvents.

\nocite{Rou:76}
\nocite{Brod:95}
\nocite{Sco:80}
\nocite{Wo:93}
\nocite{BeLeRo:97} \nocite{Hull:97}
\nocite{Bor:95} \nocite{Bor:96}
\nocite{Wo:99}

\addcontentsline{toc}{section}{References}
\newcommand{\noopsort}[1]{} \newcommand{\printfirst}[2]{#1}
  \newcommand{\singleletter}[1]{#1} \newcommand{\switchargs}[2]{#2#1}


\end{document}